# Clues on Software Engineers' Learning Styles

Luiz Fernando Capretz
Department of Electrical & Computer Engineering
University of Western Ontario
London, Ontario, N6A 5B9, CANADA
lcapretz@ eng.uwo.ca

**Abstract:** *The Myers-Briggs Type Indicator (MBTI) has proved to be a useful instrument for understanding student learning preferences and has enable comparisons of the learning preferences for various personality types. Regarding learning styles, there is no one best combination of characteristics, since each preference has its own advantages and disadvantages. Therefore, it is a fallacy to think that professors can devise a single teaching technique that would always appeal to all students at the same time. The ideas presented in this paper have been taken into account in two 4$^{th}$ year courses, named Software Requirements and Software Design in which the students develop their capstone projects. The results of this investigation may help college instructors to understanding the preferred leaning style of software engineers.*

## 1. Introduction

All of us have something to learn and to teach. Adjusting instruction to accommodate the learning styles of different types of students can increase both achievement and the enjoyment of learning. Kalsbeek [5] stated that "learning can be understood as a person's preferred approach to information processing, idea formation, and decision making; the attitude and interests that influence what is intended to in a learning situation; and a disposition to seek learning environments compatible with these personal profiles". The match or mismatch between the way that professors teach and the way that students learn may have important ramifications for levels of satisfaction and retention of students and teachers.

From their earliest years, individuals demonstrate different ways in which they learn best [6]:
- Some children prefer to get careful instructions before they began a new game or task.
- Some like to plunge in right away and learn as they go along.
- Some like to observe others playing with a toy before they try it themselves.
- Some prefer to focus by themselves.
- Some like to know all the rules and follow them.
- Some like to create their own rules and change them frequently.

Educators have been using the Myers-Briggs Type Indicator (MBTI) [6] to develop teaching methods, and to understand both individual learning styles and differences in motivation. The MBTI is neither a measure of teaching performance nor learning competence, it is only an indicator of preferences. As a rule of thumb MBTI provides insights for effective teaching and learning, and it can be usefully employed as a guide for understanding learning styles and improving teaching skills. It is this well-researched view of type theory that we would like to apply to our discussion of teaching and learning styles of engineering students [2].

To do so, we will describe several approaches to teaching, and how type is related to each approach. We feel this is the best way to improve teaching effectiveness, because it explains why teachers are sometimes pressured to teach in a way that does not suit their personality traits and how students are forced to learn in environments that do not suit their learning styles either. To understand this, it is necessary to look at a teacher's and student's preferred teaching and learning styles, as it will be discussed later in this article, but for the moment a brief description of the MBTI scales is presented below.

### 1.1 Extroversion and Introversion

Some people are oriented to a breadth-of-knowledge approach with quick action; others are oriented to a depth-of-knowledge approach reflecting on concepts and ideas. Jung calls these orientations extroversion and introversion.

### 1.2 Sensing and Intuition

Some people are attuned to the practical, hands-on, common-sense view of events, while other are more attuned to the complex interactions, theoretical implications, or new possibilities of events. These two styles of information gathering, or perception, are known as sensing and intuition, respectively.

### 1.3 Thinking and Feeling

Some people typically draw conclusions or make judgments objectively, dispassionately and analytically; others weigh the human factors or societal import, and make judgments with personal conviction as to their value. These two styles of decision-making, or judgment, are called thinking or feeling, respectively.

### 1.4 Judgment and Perception

Finally, some people prefer to collect only enough data to make decisions before setting on a direct path to a goal, and typically stay on that path. Others are finely attuned to changing situations, alert to new developments that may require a change of strategy, or even a change of goals. These two styles are called the preferences for judgment or perception, respectively.

According to MBTI, no type is better than any other and the various types are gift differing. Of course, people can and do use all eight preferences. In each of the four pairs, however, we all have one preference that is stronger than the other, one that works better for us than its complement.

The Myers-Briggs Type Indicator (MBTI) describes 16 types which result from the dynamic interplay of these four preferences – Extroversion (E) or Introversion (I), Sensing (S) or Intuition (N), Thinking (T) or Feeling (F), and Judging (J) or Perception (P). Combining the four preferences results in an overall pattern called a type. The types are denoted by the letters of preferred orientations (such as: ISTJ, ENFP, INTP, etc.) as shown in Table 1. For example, a person whose type is ENTJ has preferences for the dimensions Extroversion, Intuition, Thinking, and Judging; on the contrary, another person classified as ISFP has preference for Introversion, Sensing, Feeling, and Perception.

**Table 1**. The 16 MBTI Types

| ISTJ | ISFJ | INFJ | INTJ |
|------|------|------|------|
| ISTP | ISFP | INFP | INTP |
| ESTP | ESFP | ENFP | ENTP |
| ESTJ | ESFJ | ENFJ | ENTJ |

## 2 The Software Engineers

Software engineers have been highly stereotyped, the following diverse characteristics are accepted as part of the somewhat "unique" profile of software professionals:
- Low need for social interaction.
- High need for challenge and achievement.
- Low motivation towards management responsibilities.
- Low identification with authority.
- Low tolerance for interpersonal conflicts.
- Loyalty to profession rather than employer.
- Optimism regarding time estimates.
- Methodical approach to problem-solving.
- Interest in stable secure work.

The problem with the above personality traits is that they have typically been applied to programmers, without regard for the particular orientation of their software engineering endeavours, such as system analysis, design or maintenance. Our subjects comprise a group of software engineering students. Sixty-eight software engineers students were invited to participate in the study, and were administered the MBTI (Form G) to determine their personality types. This investigation considers students in upper level university classes. The type distribution of the software engineering students is summarized in Table 2.

**Table 2.** Type Distribution of Software Engineering Students (N = 68)

| ISTJ | ISFJ | INFJ | INTJ |
|------|------|------|------|
| N=13 | N=2  | N=1  | N=5  |
| 19.1% | 2.9% | 1.5% | 7.4% |
| ISTP | ISFP | INFP | INTP |
| N=3  | N=3  | N=2  | N=9  |
| 4.4% | 4.4% | 2.9% | 13.2% |
| ESTP | ESFP | ENFP | ENTP |
| N=8  | N=1  | N=2  | N=5  |
| 11.8% | 1.5% | 2.9% | 7.4% |
| ESTJ | ESFJ | ENFJ | ENTJ |
| N=8  | N=2  | N=1  | N=3  |
| 11.8% | 2.9% | 1.5% | 4.4% |

This study has shown that ISTJ, ESTP, ESTJ and INTP compose almost 55% of the sample, therefore, significantly over-represented, whereas ESFP, INFJ and ENFJ are all particularly underrepresented in this sample. It is also worth noticing that there are more ISTJ (19%) than any other type. This research also found more introverts (I=54%) than extroverts (E=46%) types; fairly more sensing (S=57%) than intuitive (N=43%) types; significantly more thinking (T=81%) than feeling (F=19%) types; and slightly more judging (J=54%) compared to perception (P=36%) types.

# 3 Learning Issues

Learning style is a term that refers to an individual's characteristic and consistent approach to perceiving, organizing and processing information. The idea that people have different learning styles is enticing for educators. First, it highlights the importance of learning processes, as well as teaching techniques. Second, it is an egalitarian concept because it focuses on people's strengths and weaknesses, that is, learners become *different* rather than bad, poor, average, good and excellent. Because of this, it would be naïve to expect that teachers could easily design a course and deliver a lecture to fit the learning style needs of all their students.

Myers *et al*. [6] gives a summary of findings that relate psychological types to teaching and learning styles, as following.

Extrovert students learn by talking our and interacting with others. They want faculty who encourage class discussion. Extrovert teachers give students choices and voice, are attuned to changes in students attention and comfortable with noisy classrooms. They tend to positively evaluate students who are active, energetic, and enthusiastic. On the other hand, introvert students need quiet and time for internal processing of information. They want faculty who give clear lectures. Introvert teachers structure teaching activities, are attuned to the ideas they teach and are comfortable with business-like atmosphere. They tend to positively evaluate students who are thoughtful, reflective, and introspective.

Students who prefer sensing become aware of the reality of the situation. They want faculty who give clears assignments. Sensing instructors emphasize facts, practical information, concrete skills, they usually ask for detailed and fact-oriented questions. Sensing instructors are biased to students who are factual, practical, and accurate. On the contrary, students who prefer intuition become aware of the meanings and relationships that go beyond the information that is given by focusing on the big picture. They want faculty who enforce independent thinking. Intuitive instructors emphasize concepts, implications of facts, their questions call for synthesis and meaning. They are biased to students who are conceptual, creative, and insightful.

Students who prefer thinking like to decide things objectively, based on their analysis of the logical consequences of alternatives in a detached manner. They want faculty who make logical presentations. Thinking educators talk from an objective base, they want students to focus on what he or she is doing or saying, they attend to class as a whole. They incline for students who are logical and precise, in their own work. As an opposite, students who prefer feeling like to decide things by considering what is most important to them or to the other people. They want faculty who establish personal rapport with students. Feeling educators seek dialogue, engagement, they encourage students to focus on interpersonal work; they attend to individuals or small groups. They incline for students who are pleasant to work with.

Lastly, judging type students want structure, an orderly schedule, a time frame, and closure on one topic before going on to the next; they want faculty to be organized and business-like. Judging scholars are very orderly and stick to class plan with organized lectures, they like well-arranged classroom. They tend to positively evaluate students who are task-focused, timely and organized. On the other hand, perceiving type students want flexibility; the opportunity to explore, and to follow interesting tangential information as it comes up. They want faculty to be entertaining and inspiring. Perceiving scholars are lax and less organized, they like as much activity-oriented work as possible. They tend to positively evaluate students who are easygoing.

Cooper and Miller [4] reported that the level of learning style/teaching style congruency is related to academic performance and to student evaluations of the course and instructor. Additionally, the existence of the discrepancy between students' preferences of learning in a concrete manner (S style) and faculty's penchant to teach in abstractions (N style) appears to contribute to student dissatisfaction as indicated by the course and instructor evaluations.

Blume [1] suggests that college students can improve their study habits by knowing their MBTI type and show different learning styles are associated with each preference; advice is also provided for the student whose learning style conflicts with the instructor's teaching style. Similar accounts of the relation between MBTI type and learning propensities in a software engineering course is described in Capretz [3].

# 4 Conclusions

We tend to teach, as we ourselves like to be taught; commonly we assume that our students can learn best by employing the same techniques that we used as students. However people differ significantly in the way in which they learn best; it is believed that these learning styles are related to psychological types.

Although each preference has some predictable effects on learning styles, the most significant difference is between sensing and intuition. The MBTI intuitive/sensing scale can separate intuitive students with a preference for abstract, global, and theoretical approaches from the sensing student with their preference for practical, factual, and specific approach. Sensing types can be confused by an intuitive type's use of metaphor and symbolic language, as well as by the intuitive tendency to associate from one idea to another. For sensing types, the association often leaves gaps in the development of understanding. Intuitive types, on the other hand, can become restless and inattentive with the tendency of sensing types to carefully build toward conclusions, and to include a wealth of concrete facts and specific detail.

Educators should bear in mind that everyone has a learning style that narrows their capacity as a learner. This does not mean, however, there are two classes of learners, the privileged class (learner who can overcome their limitations) and the less privileged class (learners who are not capable of using different learning styles). It is only a matter of preference, being more comfortable or not with a style. This challenges the notion that learning potential is reducible to a single dimension such as intelligence. Each learning style has its strengths and weaknesses and therefore a person locked exclusively into one style is never going to be an ideal learner.

Software engineering faculty should recognize that their classes contain all types of learners. Hence, effective instruction should try to make some appeal to each learning style for some of the time in a balanced fashion. That means incorporating activities that require reflection and occasional discussion. Challenge them with problem solving exercises involving abstraction and practice; encourage them to see the tree as well as the forest; give them the opportunity to develop a personal (feeling) touch and whenever possible, tolerate deadline flexibility to cater for the needs of the perceiving types. The type theory provides a way to deal with such issues.

In closing, the MBTI and its inferences provide a way to conceptualize a student as an organized dynamic personality which predisposes each student to certain ways of thinking, wanting, liking and acting, and gives the student a unique learning pattern. It seems reasonable to expect that a class of students encompass a variety of personality traits. Therefore, good professors should be able to broaden their teaching techniques to make teaching more effective, and so be able to reach all students at least some of the time. They should also consider varying their teaching styles on occasion to motivate and establish common ground with those few students who have traits different from their own.

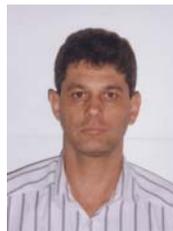

**Luiz Fernando Capretz** has vast experience in software engineering. He has worked at both technical and managerial levels, taught and done research on the engineering of software in Brazil, Argentina, England, Japan, and Canada since 1981; has published extensively and co-authored a book (Object-Oriented Software: Design and Maintenance) in the area; and has been invited to give talks all over the world. In the Faculty of Engineering at the University of Western Ontario, Dr. Capretz is currently the Director of the Software Engineering Program and teaches software design and software testing. He received his Ph.D. from the University of Newcastle upon Tyne (England), M.Sc. from INPE (Brazil), and B.Sc. from UNICAMP (Brazil); all degrees in computer science. His main research areas include: software product lines, software estimation, and human factors in software engineering. He is a senior member of IEEE, a voting member of the ACM, and a Certified MBTI practitioner.